\documentclass[doublecol]{epl2} 
\usepackage{graphicx}
\usepackage{amsmath, amssymb}
\usepackage{lmodern}
\usepackage[T1]{fontenc}
\usepackage{subcaption}
\usepackage[utf8]{inputenc}
\usepackage[english]{babel}
\usepackage{float}
\usepackage{color}
\usepackage{braket}

\title{Adaptive altruistic strategy in cyclic models during an epidemic}

\author{J. Menezes \inst{1,2} \and B. Ferreira \inst{2}   \and E. Rangel\inst{3} \and B. Moura \inst{4,5}}

\institute{   
  \inst{1} Institute for Biodiversity and Ecosystem
Dynamics, University of Amsterdam, Science Park 904, 1098 XH
Amsterdam, The Netherlands\\                 
  \inst{2} School of Science and Technology, Federal University of Rio Grande do Norte\\
59072-970, P.O. Box 1524, Natal, RN, Brazil\\
  \inst{3} Department of Computer Engineering and Automation, Federal University of Rio Grande do Norte,  Av. Senador Salgado Filho 300, Natal, 59078-970, Brazil\\
  \inst{4} Edmond and Lily Safra International Institute of Neuroscience, Santos Dumont Institute,
Av Santos Dumont 1560, 59280-000, Macaiba, RN, Brazil\\                 
  \inst{5} Department of Biomedical Engineering, Federal University of Rio Grande do Norte,
 Av. Senador Salgado Filho 300, Lagoa Nova, 59078-970, Natal, RN, Brazil\\
 }


\abstract{
We investigate a cyclic game system where organisms face an epidemic beyond being threatened by natural enemies. As a survival strategy, individuals of one out of the species usually safeguard themselves by approaching the enemies of their enemies and performing social distancing to escape contamination when an outbreak affects the neighbourhood.  
We simulate how the survival movement strategy to local epidemic surges must adapt if a pathogen mutation makes the disease deadlier. 
We study the spatial distribution of local outbreaks and observe the influence of disease mortality on individuals' spatial organisation.
We show that adapting the survival movement strategy for a high mortality disease demands an altruistic behaviour of the organisms since their death risk increases. Despite weakening the disease transmission chain, which benefits the species, abandoning
refuges provided by safeguarding social interaction
increases the vulnerability to being eliminated in the cyclic game.
Considering that not all individuals exhibit altruism, we find the relative growth in the species density as a function of the proportion of individuals behaving altruistically.
Our results may be helpful for biologists and data scientists to understand how adaptive altruistic processes can affect population dynamics in complex systems.}

\begin{document}

\maketitle

\section{Introduction}
Individuals' spatial organisation plays a central role in the formation and stability of ecosystems \cite{ecology}. Researchers have shown that three strains of bacteria \textit{Escherichia coli} -
whose cyclic dominance is described by the rock-paper-scissors game rules - survive only if the interactions are local, leading to the formation of departed spatial domains \cite{bacteria,Coli,Allelopathy}. This remarkable experiment highlights the central role of space in biodiversity maintenance, with mobility being crucial for promoting species coexistence \cite{Reichenbach-N-448-1046
}.
Furthermore, adapting to local environmental changes allows the organisms 
to increase their fitness by moving to regions with abundant natural resources \cite{climatechange,adap2,foraging,BUCHHOLZ2007401}. The purposeful movement also has been crucial for organisms to search for refuges when threatened by competitors, making alliances with other species to protect themselves against natural enemies\cite{adaptive1,adaptive2,Dispersal,BENHAMOU1989375,Causes,MovementProfitable,howdirectional,Agg,Adap,AdapII,hamming,expanding}. 
The adaptive animal behaviour has also inspired engineers to develop tools that improve the robots' movement \cite{animats}.

Moreover, survival movement tactics have been reported in systems infested by plagues like pathogens causing infectious diseases, which increases organisms' death risk \cite{epidemicbook,epidemicprocess,COVID,disease5}. The ability to scan the environment and interpret the sensory information can be decisive to organisms identifying dangers, thus isolating themselves, diminishing the disease contamination probability \cite{Odour,socialdist,soc}. In many biological systems, the defence strategy is self-destructive for the individuals but beneficial to the species \cite{alt1}. Exhibiting altruism causes the loss of individual fitness but rewards the species whose population grows due to the increase in the collective fitness \cite{alt1,alt2}.

In this letter, we study a cyclic game system of five species where
organisms face a disease transmitted person-to-person, indistinctly affecting all species \cite{social1,disease3}. It has been shown that epidemic spreading may promote biodiversity in cyclic models \cite{ref1-1}.
Following the model introduced in Ref.~\cite{adaptive}, each organism can perceive the neighbourhood, distinguishing between healthy and sick individuals. If the density of ill organisms is below a tolerable threshold, the individual ignores the local surge and searches for protection against elimination in the cyclic game. It has been shown in Ref.~\cite{Moura} that the Safeguard strategy, which consists in moving in the direction with more enemies of the enemies, provides satisfactory protection, diminishing the individual's selection risk, leading the species to occupy the most significant fraction of the territory \cite{Moura,enenyenemy}.
But, if the local density of sick organisms is high, the social Safeguard tactic may lead the organism to approach viral vectors \cite{socialdist,soc}. In this case, the execution of the Social Distancing strategy is triggered, with the organism moving to the direction with the highest density of empty spaces \cite{doi:10.1126/science.abc8881,combination,adaptive}.

We consider that a pathogen mutation alters the disease virulence by increasing the mortality rate \cite{mut1,mut2}. We address the questions: i) 
how does high disease mortality affect the local outbreaks?; ii) what are the effects on the spatial organisms' organisation?; iii) how must individuals adjust the social distancing trigger to maximise the species density?; iv) how does the adaptation of the social distancing trigger to a deadlier disease impact individuals' death risk?; v) how does organisms' altruistic behaviour influences the results of the survival movement strategy?
\begin{figure}
\centering
\includegraphics[width=40mm]{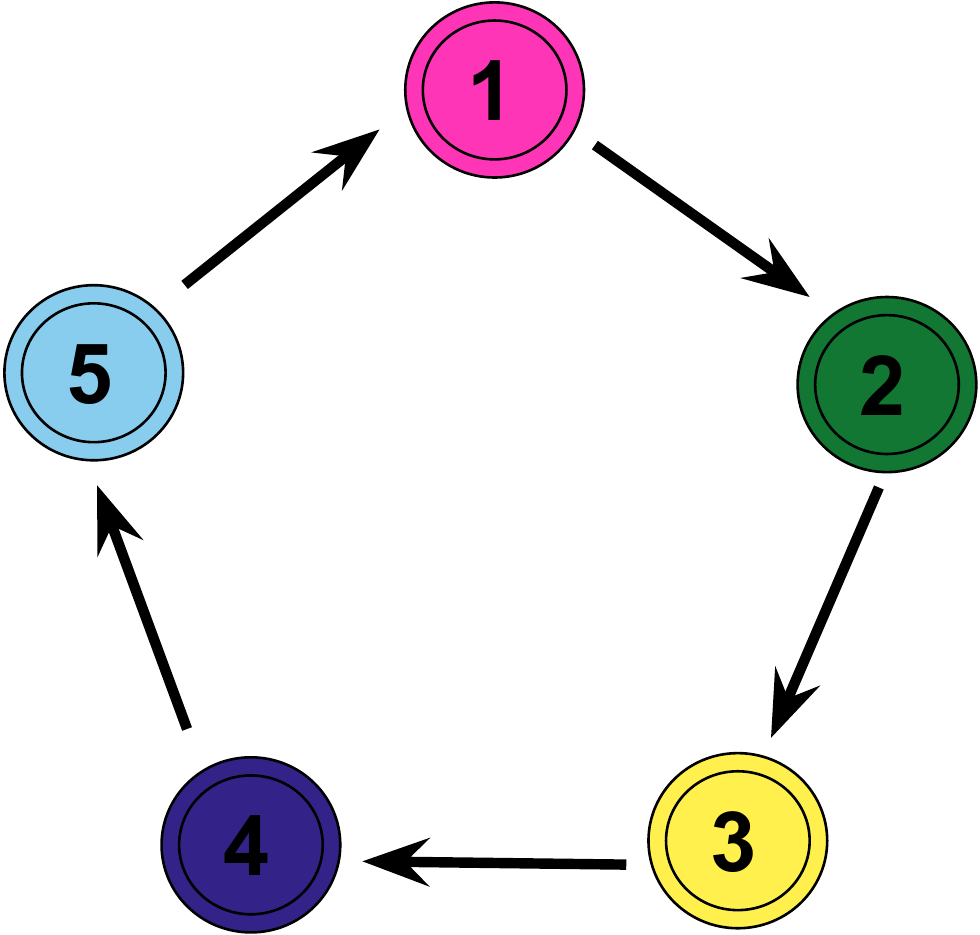}
\caption{Illustration of the generalised rock-paper-scissors model with $5$ species. Selection interactions are represented by arrows indicating the dominance of organisms of species $i$ over individuals of species $i+1$.}
\label{fig1}
\end{figure}
\begin{figure*}[t]
\centering
        \begin{subfigure}{.23\textwidth}
        \centering
        \includegraphics[width=35mm]{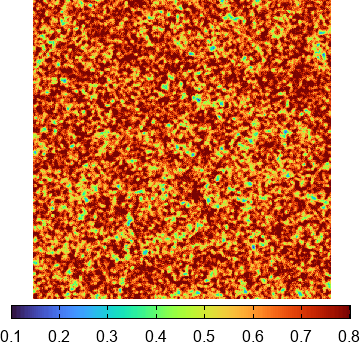}
        \caption{}\label{fig2a}
    \end{subfigure} %
    \begin{subfigure}{.23\textwidth}
        \centering
        \includegraphics[width=35mm]{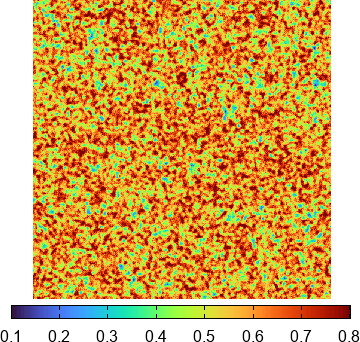}
        \caption{}\label{fig2b}
    \end{subfigure} %
       \begin{subfigure}{.23\textwidth}
        \centering
        \includegraphics[width=35mm]{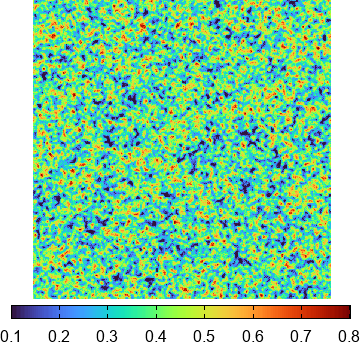}
        \caption{}\label{fig2c}
    \end{subfigure} %
           \begin{subfigure}{.23\textwidth}
        \centering
        \includegraphics[width=35mm]{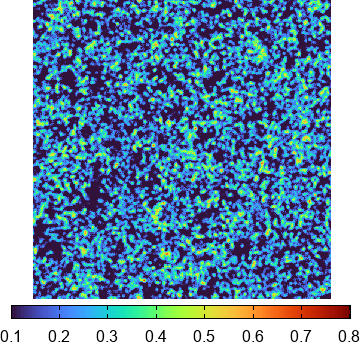}
        \caption{}\label{fig2d}
    \end{subfigure} %
\caption{Spatial distribution of the local disease outbreaks captured from simulations in lattices with $500^2$ grid points for various disease mortality rates.
Figures ~\ref{fig2a}, ~\ref{fig2b}, ~\ref{fig2c}, and ~\ref{fig2d} show the results for $\mu=0.1$, $\mu=0.2$, $\mu=0.5$, and $\mu=0.8$ at $t=5000$ generations. The color bars indicate the local density of sick individuals.
}
  \label{fig2}
\end{figure*}
\section{The model}
\label{sec2}
Our cyclic model comprises five species that outcompete one another according to the generalised rock-paper-scissors model, illustrated in Fig.~\ref{fig1}. Organisms of species $i$ kill individuals of $i+1$, with $i=1,2,3,4,5$; one assumes the cyclic identification $i=i+5\,\alpha$, where $\alpha$ is an integer.
This is the simplest generalisation of the spatial rock-paper-scissors models, where spiral patterns arise from random initial conditions \cite{ref1-sp}.

Besides the risk of elimination during an eventual attack in the cyclic game, organisms may also die due to complications of a contagious disease. The epidemic spreads through the system, passing from one organism to another by social interaction; all individuals are susceptible to contamination, irrespective of the species. Once infected, organisms become viral vectors, passing the virus to immediate neighbours before being cured or dying because of disease complications. Cured organisms do not gain immunity, thus being vulnerable to being reinfected.

We study a system where organisms of one out of the species perform an evolutionary strategy to minimise the death risk. Following the model introduced in Ref.~\cite{adaptive}, we 
simulate the movement tactic:

\begin{itemize}
\item
Usually, individuals move in the direction with the highest density of enemies of their enemies, thus, benefiting from the protection against death in the cyclic game. The goal of the Safeguard strategy is to diminish the probability of being selected.
\item
If a disease surge reaches the neighbourhood, agglomerating with enemies of the enemies becomes hazardous; therefore, it is time for self-isolation.
The Social Distancing strategy aims to reduce the chances of disease contamination.
\end{itemize}

The self-preservation tactic is locally adaptive: each organism autonomously decides the more appropriate movement strategy at each instant. This is possible because individuals can scan the neighbourhood and detect the presence of sick organisms. A local surge stimulates individuals to move in the direction with more empty spaces if the density of viral vectors is higher than a tolerable threshold: the social distancing trigger. 
Organisms' adaptation to changes in the disease virulence is mainly altruistic, not aiming to maximise the protection against disease infection and natural enemies but for the overall benefit of the species. Namely, individuals accommodate the social distancing trigger to reduce the threats, guaranteeing the predominance in the cyclic competition for space.

Our numerical realisations are performed in square lattices with periodic boundary conditions. The total number of grid points is $\mathcal{N}$, the maximum number of organisms present in the system (each grid point contains at most one individual). We use 
the May-Leonard implementation, where the total number of individuals is not conserved \cite{leonard}. 
To build the initial conditions, we allocate the same number of organisms for every species at random grid points. Defining the total number of individuals $I_i$, we write
the initial number of individuals as $I_i\,=\,\mathcal{N}/5$, with $i=1,2,3,4,5$ - there are no empty spaces in the initial conditions.
To distinguish between sick and healthy individuals of species $i$, we introduce the notation $h_i$ and $s_i$ (the notation $i$ means any organism, independent of its health conditions) \cite{combination, adaptive}.

Following the von Neumann neighbourhood, an organism can interact with one of its four immediate neighbours according to the rules:
a) Selection: $ i\ j \to i\ \otimes\,$, with $ j = i+1$, where $\otimes$ means an empty space; b) Reproduction: $ i\ \otimes \to i\ i\,$; c) Mobility: $ i\ \odot \to \odot\ i\,$, where $\odot$ means either an individual of any species or an empty site;
d) Infection: $ s_i\ h_j \to s_i\ s_j\,$, with $i, j=1,2,3,4,5$;
e) Cure: $ s_i \to h_i\,$; 
f) Death: $ s_i \to \otimes\,$. Empty spaces are created when individuals are selected, or sick organisms die because of the disease severity. 

The occurrence of a given interaction in the stochastic simulations is proportional to the 
set of real parameters: $\mathcal{S}$ (selection rate), $\mathcal{R}$ (reproduction rate), $\mathcal{M}$ (mobility rate), $\kappa$ (infection rate), $\mu$ (mortality rate), and $\omega$ (cure rate); the parameters are the same for all organisms of every species.
The implementation follows three steps: i) an active organism is randomly chosen among all individuals in the lattice;
ii) an interaction is randomly chosen according to the set of rates;
iii) one of the four nearest neighbours to suffer the action; the exception is the directional movement strategy, where the direction depends on the specific survival strategy.

At each time step, one interaction modifies the organisms' distribution, which may alter the fraction of lattice occupied by each species. To quantify the population dynamics, we calculate the densities of individuals of species $i$ at time $t$, defined as 
$\rho_i = I_i/\mathcal{N}$, with $i=1,2,3,4,5$. Our time unit 
is called generation, which defines the necessary time for $\mathcal{N}$ timesteps to occur. 
\begin{figure*}
\centering
    \begin{subfigure}{.23\textwidth}
    \centering
    \includegraphics[width=35mm]{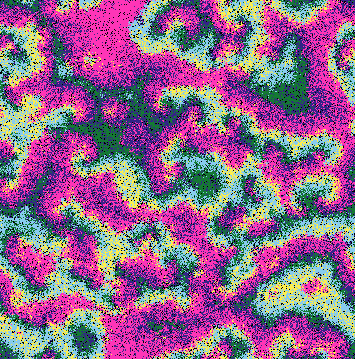}
    \caption{}\label{fig3a}
  \end{subfigure} %
  \begin{subfigure}{.23\textwidth}
    \centering
    \includegraphics[width=35mm]{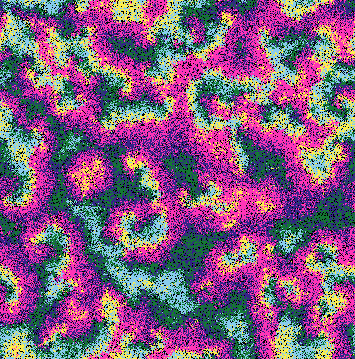}
    \caption{}\label{fig3b}
  \end{subfigure} %
    \begin{subfigure}{.23\textwidth}
    \centering
    \includegraphics[width=35mm]{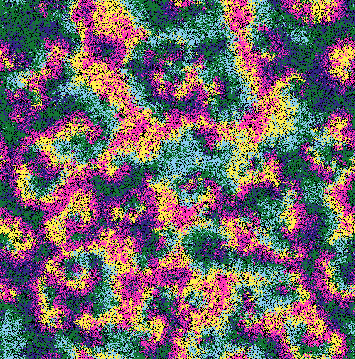}
    \caption{}\label{fig3c}
  \end{subfigure} %
      \begin{subfigure}{.23\textwidth}
    \centering
    \includegraphics[width=35mm]{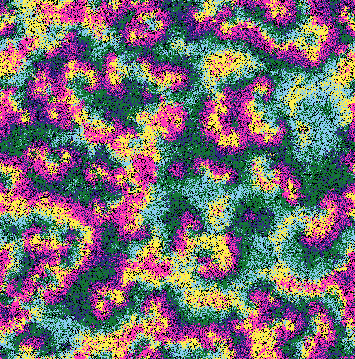}
    \caption{}\label{fig3d}
  \end{subfigure} %
\caption{Spatial patterns captured from simulations shown in Fig.~\ref{fig2}. The colours follow the scheme in Fig.~\ref{fig1}; empty spaces appear in white dots. The snapshots in Figs.~\ref{fig3a}, \ref{fig3b}, \ref{fig3c}, and \ref{fig3d} shows the spatial organisation after $t=1500$ generations for $\mu_A=0.1$, $\mu_B=0.2$, $\mu_C=0.5$, and $\mu_D=0.8$, respectively.
}
 \label{fig3}
\end{figure*}
\begin{figure}
   \centering
  \includegraphics[width=85mm]{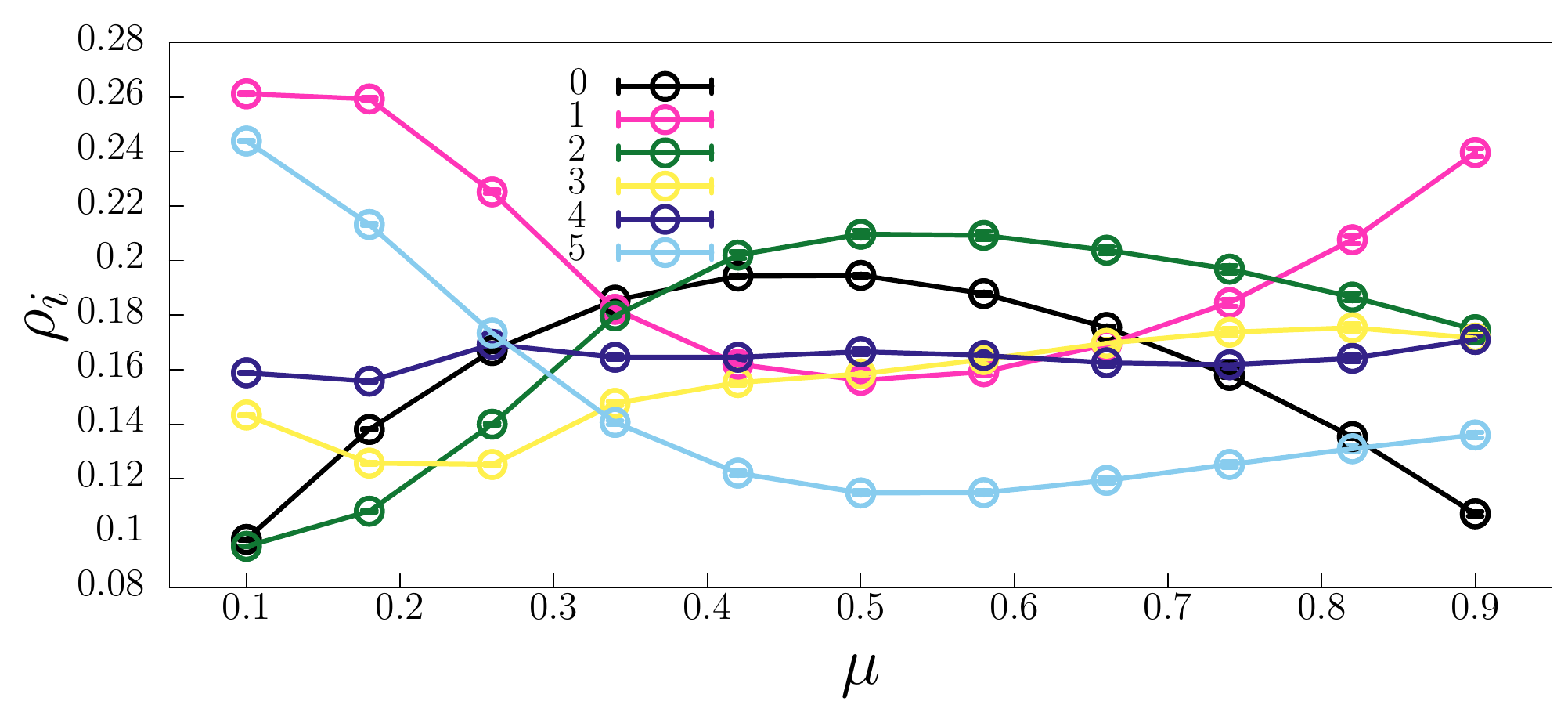}
  \caption{Species densities as a function of the mortality disease. The 
colours refer to the species following the scheme in Fig.~\ref{fig1}. The black line depicts the density of empty spaces; error bars show the standard deviation.}
 \label{fig4}
\end{figure}
\begin{figure}
   \centering
  \includegraphics[width=85mm]{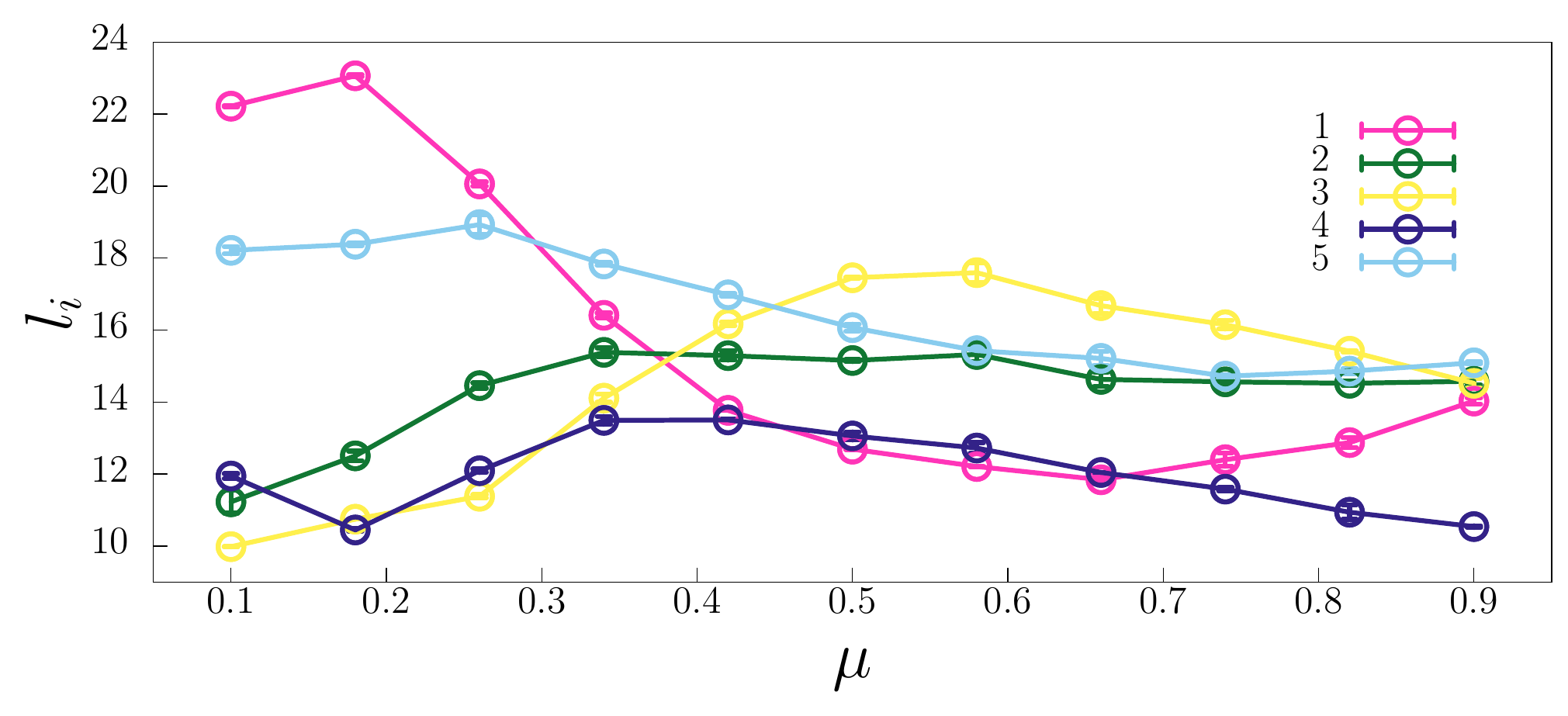}
  \caption{Characteristic length scale of the spatial agglomerations of species $i$ as a function of $\mu$. The error bars show the standard deviation.}
 \label{fig5}
\end{figure}

The implementation of the locally adaptive survival movement strategy follows the steps \cite{Moura,combination}:
\begin{enumerate} 
\item 
We define the perception radius $R$, measured in lattice spacing, as the maximum distance an individual can scan the environmental cues to use the information to decide what strategic movement is suitable at the moment, Safeguard or Social Distancing.
\item
We implement a circular area of radius $R$ centred at the active individual, defining the total grid points the organism can analyse.
\item
We define a real parameter $\beta$, with $0\,\leq\,\beta\,\leq\,1$, to define the social distancing trigger parameter: a threshold assumed by the organisms as the minimum local density of sick individuals to perform the social distancing strategy.
\item
We calculate the density of sick organisms within the 
disc of radius $R$ surrounding the active individual. In case of the local density is below the threshold $\beta$, the organism decides to perform the Safeguard strategy; otherwise, the Social Distancing tactic is chosen.
\item
We separate the organism's perception area
into four circular sectors in the directions of the nearest neighbours. Thus, we count the number of empty spaces and organisms within each circular sector; individuals on the borders are assumed to be in both circular sectors.
\item
We define the movement direction according to the organism's strategy choice: the direction with the more significant number of organisms of species $i-2$ ($h_{i-2}+s_{i-2}$) for Safeguard or the direction with more empty spaces for Social Distancing. If more than one direction is equally attractive, a draw is done.
\end{enumerate}

Throughout this letter, all results were obtained from simulations in square lattices with $500^2$ sites, running for a timespan of $5000$ generations, for $R=3$. Initially, the proportion of sick organisms is $1\%$.
Furthermore, we study the variation in the disease virulence caused by virus mutation, considering a range of $\mu$.
Therefore, we define the interaction rates as follows: 
$\mathcal{S} = \mathcal{R}=\mathcal{M}=1$, $\kappa=2$, $\omega=0.2$, with $\mu$ being set according to the experiment, as described in the Figures' captions. However, we confirmed that our conclusions are valid for other sets of interaction rates by repeating the simulations for other parameters.
The statistical analyses whose results appear in Figs.~\ref{fig4} to ~\ref{fig6} were realised by performing a series of $100$ simulations, starting from different initial conditions; in each case, the standard deviation is represented by error bars.
\section{Spatial Patterns}

We first observe the effects of disease mortality in the spatial patterns by running a single simulation for various values of $\mu$, considering $R=3$ and $\beta=0.6$. We name the realisation as Simulation A, B, C, and D, for $\mu_A=0.1$, $\mu_B=0.2$, $\mu_C=0.5$, and $\mu_D=0.8$, respectively. 

Figures \ref{fig2a}, \ref{fig2b}, \ref{fig2c}, and \ref{fig2d} shows the local density of sick individuals at each grid point at the end of the simulations A, B, C, and D, respectively. 
The colour bar displays the magnitude of the local density of sick individuals at each lattice site. The non-homogeneous spatial distribution of ill organisms indicates that epidemic outbreaks reach different parts of the grid differently. For higher mortality, infected organisms are more likely to die, making the disease surges less intense. At
the moment the snapshot in Fig.~\ref{fig2a} was captured, most of the individuals of species $1$ were executing the Social Distancing strategy since the density of sick organisms is higher than $60\%$ at most grid points. The proportion of individuals performing the Social Distancing decreases as $\mu$ increases, as shown in Figs.~\ref{fig2b} to \ref{fig2d}. Thus, for a fixed social distancing trigger, 
the frequency of execution of the Social Distancing tactic decreases with $\mu$ (the Safeguard strategy is more used as $\mu$ grows).

Figures \ref{fig3a}, \ref{fig3b}, \ref{fig3c}, and \ref{fig3d} show the final spatial organisms' organisation at the end of the realisations for Simulations A, B. C, and D, respectively. The entire simulation is shown in videos https://youtu.be/4MA3hI7XGug ($\mu_A=0.1$), https://youtu.be/NTazzkgVcmA ($\mu_A=0.2$), https://youtu.be/F9TrCKXxmVg ($\mu_A=0.5$), https://youtu.be/GKEgSlFwuJs ($\mu_A=0.8$), respectively.
Individuals are shown in the respective species colour appearing in Fig.~\ref{fig1}: pink, green, yellow, purple, and blue stand for individuals of species $1$, $2$, $3$, $4$, and $5$, respectively; black dots represent empty spaces. 

Organisms of species $1$ use the information about the local density of sick individuals to decide if the Social Distancing strategy should be performed. Therefore, as in Figs.~\ref{fig2a}, the average density of ill organisms is higher, the Social Distancing is more frequently triggered; thus, organisms significantly benefit from protection against disease contamination. Consequently, the areas occupied by organisms of species $1$ (pink regions) are larger than the regions occupied by the other species in Fig. \ref{fig3a}, which represents an advantage in the cyclic game \cite{uneven}. 

As $\mu$ increases, the probability of the local density of ill individuals being more than $60\%$ of the neighbourhood decreases. Since $\beta=0.6$, reducing the frequency of execution of the Social Distancing tactic. This means that part of the individuals of species $1$ goes in the direction with the highest proportion of empty spaces, while the others approach individuals of species $4$. Because of this, the average size of the pink regions decreases, as shown in Figs.~\ref{fig3b} and ~\ref{fig3c}. Nevertheless, as Fig. ~\ref{fig2d} shows that for $\mu=0.8$, the number of sick individuals is significantly reduced; in this case, almost
all organisms of species $1$ perform the Safeguard strategy, thus 
the average size of the spatial domains inhabited by individuals of species $1$ grows, as observed in Fig.~\ref{fig3d}.

\section{Spatial domains' characteristic length scale}

Now, we investigate how the scale of typical spatial domains of individuals of the same species depends on the disease mortality.
For this purpose, we first find the spatial autocorrelation function in terms of radial coordinate $r$ for individuals of each species, 
$C_i(r)$, with $i=1,2,3,4,5$ \cite{combination}.
We introduce the function $\phi_i(\vec{r})$ that identifies the position $\vec{r}$ in the lattice occupied by individuals of species $i$. Using the mean value $\langle\phi_i\rangle$ to find the Fourier transform
\begin{equation}
\varphi_i(\vec{\kappa}) = \mathcal{F}\,\{\phi_i(\vec{r})-\langle\phi_i\rangle\}, 
\end{equation}
and the spectral densities $S_i(\vec{k}) = \sum_{k_x, k_y}\,\varphi_i(\vec{\kappa})$.

We then calculate the normalised inverse Fourier transform to find the autocorrelation function for species $i$ as
\begin{equation}
C_i(\vec{r}') = \frac{\mathcal{F}^{-1}\{S_i(\vec{k})\}}{C(0)},
\end{equation}
which can be written as a function of $r$ as
\begin{equation}
C_i(r') = \sum_{|\vec{r}'|=x+y} \frac{C_i(\vec{r}')}{min\left[2N-(x+y+1), (x+y+1)\right]}.
\end{equation}
The characteristic length scale of the spatial domains of species $i$ as $l_i$, with $i=1,2,3,4,5$, is found by assuming the 
threshold $C_i(l_i)=0.15$.
We calculate $l_i$ using a sets of $100$ simulations for the range disease mortality: $0.1 \leq \mu \leq 0.9$. The social distancing trigger and the individuals' perception radius are the same as in Fig.\ref{fig4} - the colours represent the species, according to the scheme in Fig.\ref{fig1}.

The outcomes depicted in Fig.\ref{fig5} show that the mean value of characteristic length scale $l_i$, with $i=1,2,3,4,5$. Organisms of species $1$ form the spatial domains with the longest characteristic length scale for $\mu=0.18$, with species $5$ occupying the second largest areas. 
As $\mu$ grows, the pink line shows a decrease in the average size for $ \mu\leq 0.5$, which is reverted for $ \mu> 0.66$. This happens because the benefits of the adaptive survival movement strategy accentuate as the average number of viral vectors surrounding the individuals of species $1$ drops, which leads the Safeguard strategy to be more often executed. 

For high mortality disease, the intensity of the local surges loses relevance, with individuals practically executing the Safeguard strategy. In this case, all organisms of species $1$ join groups of individuals of species $4$; thus, individuals of species $4$ are less spatially correlated, as observed in Fig.\ref{fig3d} for $\mu=0.8$. This is the reason the purple line in Fig.\ref{fig5} decreases significantly for $\mu \geq 0.66$.

\section{Impact of disease mortality on population dynamics}
\label{sec6}

We explore the effects of variation in disease mortality 
on the population dynamics by performing a series of $100$ simulations, starting from different initial conditions, for $\beta=0.6$. The outcomes depicted in Fig. \ref{fig4}
shows the mean densities of species $i$, for $0.1 \leq \mu \leq 0.9$. The colours represent the species following the scheme in Fig. \ref{fig1}, while the black line depicts the density of empty spaces.
\begin{itemize}
\item
For $\mu=0.1$, fewer individuals die due to complications from the disease; therefore, the density of empty spaces is minimum.
As $\mu$ grows, deaths of sick individuals become more frequent, producing more empty spaces. However, for $\mu>0.5$, the probability of sick individuals dying before transmitting the disease is high, thus weakening the disease spread. For this reason, the density of empty spaces declines for high mortality diseases.
\item
For low disease mortality, the social distancing trigger $\beta=0.6$ allows the organisms to properly control the adaptive survival movement tactic to benefit the species. Therefore, species $1$ predominates in the cyclic game, occupying the most significant fraction of the lattice. 
However, as $\mu$ grows, the fixed social distancing trigger loses the efficiency in detecting the presence of local surges since the average density of sick individuals decreases. Because of this, $\rho_1$ drops, reaching the minimum value for $\mu=0.5$. 
For $\mu>0.5$, the number of sick individuals surrounding the individuals of species $1$ diminishes significantly; thus, the disease contamination risk is reduced, allowing the population of species $1$ to rise again, prevailing over other species for $\mu \geq 0.4$.
\end{itemize}

\begin{figure}[t]
\centering
       \begin{subfigure}{.49\textwidth}
        \centering
        \includegraphics[width=75mm]{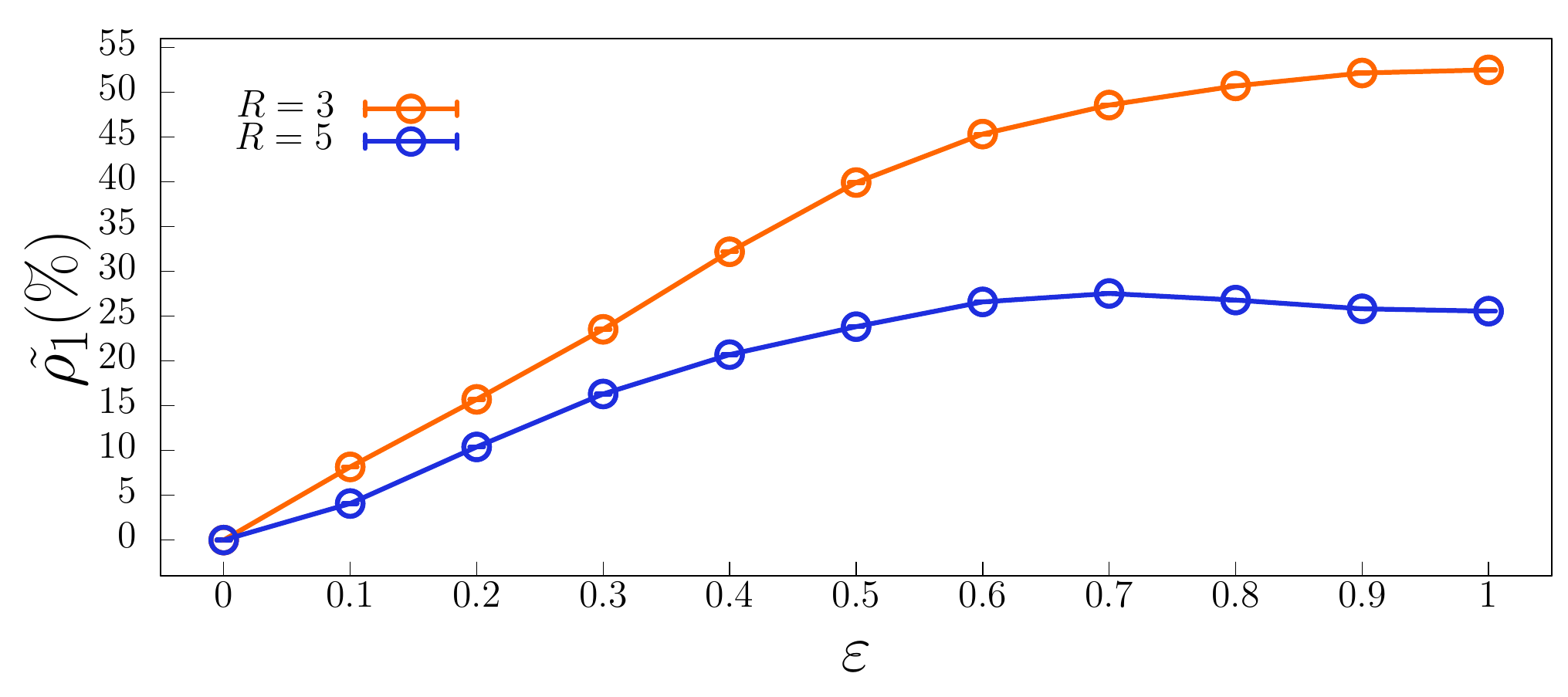}
        \caption{}\label{fig6a}
    \end{subfigure} %
        \begin{subfigure}{.49\textwidth}
        \centering
        \includegraphics[width=75mm]{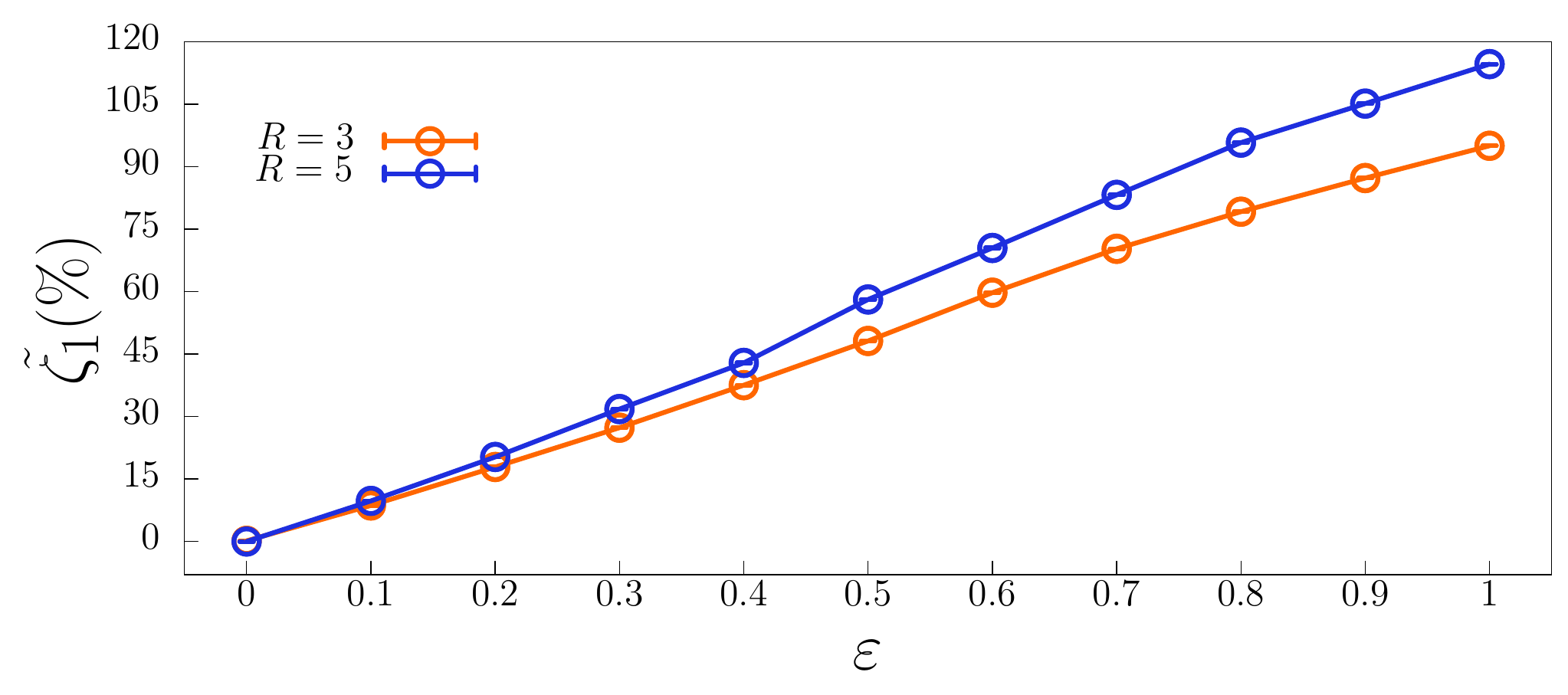}
        \caption{}\label{fig6b}
    \end{subfigure} %
       \begin{subfigure}{.49\textwidth}
        \centering
        \includegraphics[width=75mm]{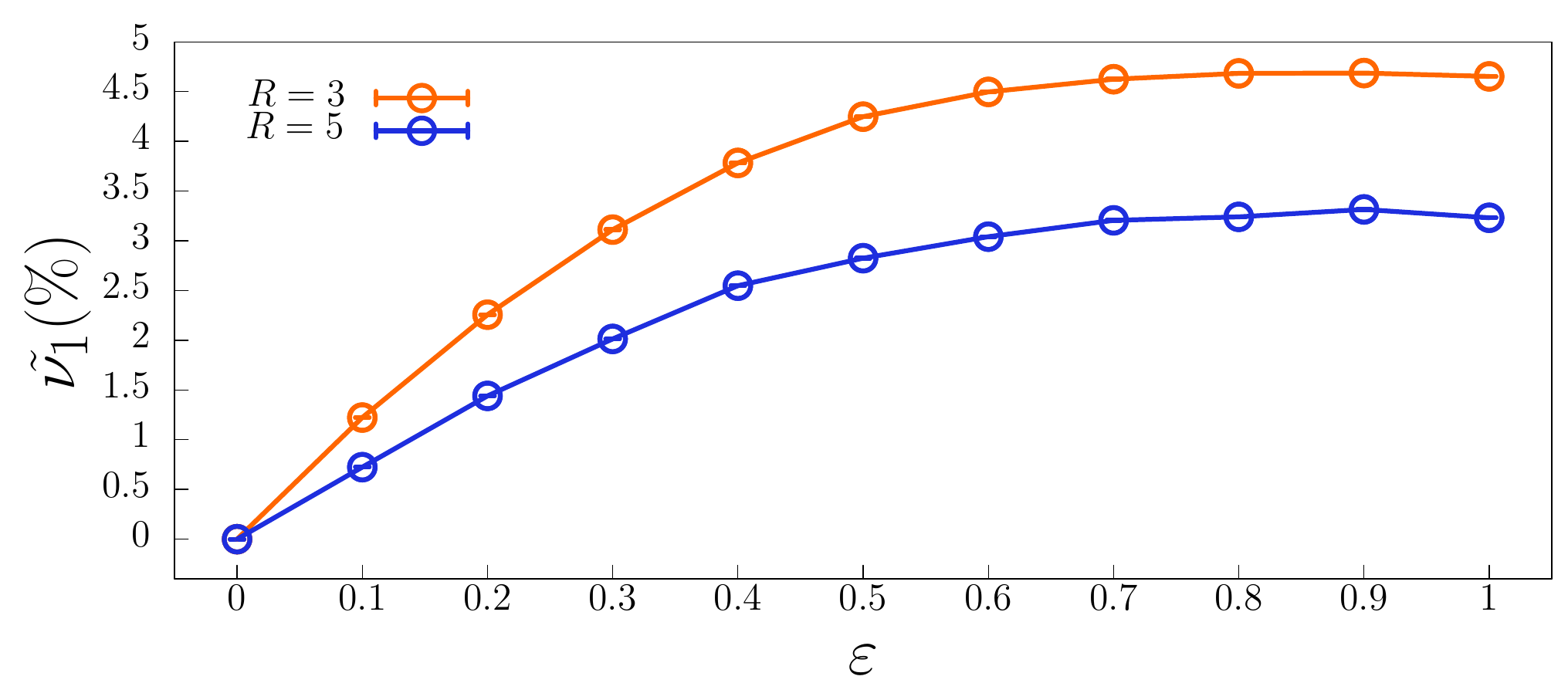}
        \caption{}\label{fig6c}
    \end{subfigure} %
    \caption{Relative variations in spatial density (figure a), selection risk (figure b), and death risk (figure c) for species $1$. Orange and blue lines depict the results for $R=3$ and $R=5$, respectively, averaged from sets of $100$ simulations; the error bars show the standard deviation.}
  \label{fig6}
\end{figure}

\section{The role of the organisms' altruistic behaviour}
\label{sec6}
Finally, we investigate the role of altruism in adapting the survival strategy when the pathogen mutation makes the disease deadlier. 
The adaptation aims to guarantee the maximum growth in the species population, leading to the predominance in the spatial game. This happens because once the disease becomes deadlier, the efficiency of the local strategy is compromised, thus adjusting the social distancing trigger is necessary to maximise the fraction of the lattice occupied by the species.
We define the altruistic factor, $\varepsilon$, a real parameter with $0 \leq \varepsilon \leq 1$, which represents the proportion of organisms of species $1$ acting to benefit the species, even if this action increases the individual death risk. For $\varepsilon=0$, no individual exhibits 
altruism, while $\varepsilon=1$, all organisms behave altruistically.

In this study, we introduce the organisms' selection and death risks, $\zeta_1$ and $\nu_1$, as follows: i) we count the total number of individuals of species $1$ when each generation begins; 
ii) we calculate how many individuals of species $1$ are killed by individuals of species $5$ during the generation;
iii) we calculate how many individuals of species $1$ die either eliminated by individuals of species $5$ or due to complications from the disease during the generation. The selection risk is the ratio between the number of selected and the initial total number of organisms, while  
the death risk is defined as the ratio between the total number of dead individuals and the initial total number of organisms.

We ran sets of realisations for the two cases: $\mu=0.12$ (low mortality disease) and $\mu=0.52$ (high mortality disease) 
for $R=3$ and $R=5$. 
For each $\mu$, $100$ different initial conditions were assumed for $0 \leq \beta \leq 0.96$, in intervals of $\Delta \beta = 0.06$.
We then define the optimum social distancing trigger $\beta^{\star}$, as the threshold that optimises the benefits of the locally adaptive survival strategy resulting in the maximum density of species $1$. 

The maximum species density is achieved if organisms assume the following optimum social distancing triggers: 
i) for $R=3$: $\beta^{\star}(\mu=0.12)=0.62$ and $\beta^{\star}(\mu=0.52)=0.33$; ii) for $R=5$: $\beta^{\star}(\mu=0.12)=0.65$ and $\beta^{\star}(\mu=0.52)=0.36$.  This happens because if a virus mutation makes the illness deadlier, the transmission chain is weakened since more sick organisms die before passing the disease. As the average local density of ill individuals drops, the social distancing alert is rarely triggered, making it difficult for organisms to protect themselves against virus contamination. Therefore, it is more advantageous if organisms lower $\beta$, thus improving the awareness of the presence of a local surge without losing the help of guards against selection. 
We then calculate the relative variations of species density and 
risks, given respectively, by
$\tilde{\rho_1}=(\rho_1-\rho^0_1)/\rho^0_1$, $\tilde{\zeta_1}=(\zeta_1 -\zeta^0_1)/\zeta^0_1$, 
$\tilde{\nu_1}=(\nu_1 -\nu^0_1)/\nu^0_1$,
where $\rho^0_1$, $\zeta^0_1$, and $\nu^0_1$ are the species densities and individuals' risks for $\varepsilon=0$.

Figures \ref{fig6a}, \ref{fig6b}, and \ref{fig6c} 
show $\tilde{\rho_1}$, $\tilde{\zeta_1}$, and $\tilde{\nu_1}$, as functions of the altruism factor $\varepsilon$. The results reveal that as the proportion of individuals accommodating the social distancing trigger 
when the disease becomes deadlier grows, the fraction of the territory occupied by species $1$ increases.
However, this occurs are the cost of individuals' exposure to being killed by individuals of species $5$, as depicted in Fig.~\ref{fig6a}: the more significant the fraction of selfless individuals, the higher the organisms' selection risk. Since the number of deaths caused by the disease drops as $\varepsilon$ grows, the maximum relative increase in the organisms' death risk occurs when $80\%$ of organisms behave altruistically, with a slight drop for $\varepsilon=1.0$. Finally, the outcomes show that for larger $R$, the threshold adaptation to the environmental changes becomes less local, thus reducing the accuracy of the decision-making and the substantial benefits of population growth.

\section{Discussion and Conclusions}
\label{sec7}
We investigate the generalised rock-paper-scissors model where organisms of one out of the species protect themselves by approaching the enemies of their enemies. However, a disease spreading person-to-person makes the social self-preservation movement dangerous, increasing the individuals' contamination risk. Therefore, every time the local density of sick individuals surpasses a tolerable threshold, the individual performs social distancing, moving in the direction with the highest density of empty spaces. When isolating socially, the individual gains protection from possible contamination but loses refuge against enemies in the spatial game. For this reason, social distancing is triggered only if the individual concludes that the proportion of sick individuals surrounding it is greater than a tolerable fixed value. The social distancing trigger cannot be so high as to prevent the execution of the Social Distancing tactic (increasing susceptibility to infection), nor so low as to abandon the Safeguard strategy (increasing vulnerability to being killed by selection). We consider that each organism is autonomous to choose between executing the Safeguard or Social Distancing strategies to produce the maximum rise 
of the species population.

Our stochastic simulations show how the local density of sick individuals varies when a virus mutation produces a high mortality disease. We verified that the average density of viral vectors decreases as the disease becomes deadlier since more sick individuals die before transmitting the virus. This affects the organism's perception of local surges, which prevents the execution of the Social Distancing strategy and, consequently, provokes the decline of the species population.

Using the results from a massive set of simulations, we discovered how the social distancing trigger must be adapted to maximise the species population. As deadlier the disease becomes, the lower the threshold to a tolerable local density of sick individuals should be, thus allowing organisms to perceive the presence of local outbreaks when the local density of viral vectors is low. However, although lowering the social distancing trigger benefits the species with population growth, it causes a rise in individuals' death. This happens because to weaken the disease transmission chain, the organisms lose the protection against enemies, 
becoming more vulnerable to being eliminated in the cyclic spatial game.
This leads to the conclusion that the species' predominance in the cyclic game with high mortality disease is reached at the cost of the sacrifices of the individuals.
Furthermore, the outcomes show that the maximum relative increase in the death risk occurs not if all individuals behave altruistically but if $90\%$ of the population adapts the social distancing trigger to the collective benefit.

\acknowledgments
We thank CNPq, ECT, Fapern, and IBED for financial and technical support.

\end{document}